\documentclass[twocolumn,showpacs,secnumarabic,amssymb,nobibnotes,aps,prl]{revtex4}

\usepackage{amsmath}
\usepackage{latexsym}
\usepackage{float}
\usepackage{amssymb}
\usepackage{graphicx}
\usepackage{textcomp}
\usepackage{hyperref}
\begin{document}

\title{Kinetics of Phase Separation in Fluids: A Molecular Dynamics Study}
\author{Shaista Ahmad$^{1,2}$, Subir K. Das$^{2,*}$ and Sanjay Puri$^{1}$}
\affiliation{$^1$School of Physical Sciences, Jawaharlal Nehru University, New Delhi 110067, India\\
$^2$Theoretical Sciences Unit, Jawaharlal Nehru Centre for Advanced Scientific Research, Jakkur P.O., Bangalore 560064, India}

\date{\today}

\begin{abstract}
We present results from extensive 3-$d$ molecular dynamics (MD) simulations of phase separation kinetics in fluids. A coarse-graining procedure is used to obtain state-of-the-art MD results. We observe an extended period of temporally linear growth in the viscous hydrodynamic regime. The morphological similarity of coarsening in fluids and solids is also quantified. The velocity field is characterized by the presence of monopole-like defects, which yield a generalized Porod tail in the corresponding structure factor.
\end{abstract}
\maketitle

The nonequilibrium evolution of a phase-separating binary mixture, A+B, is a complex nonlinear process \cite{Bray}. This problem has attracted much research interest both computationally \cite{Das1} and experimentally \cite{Blondiaux}. The growth of A-rich and B-rich domains during phase separation is a scaling phenomenon. The two-point equal-time correlation function, $C_{\psi\psi}(r,t)$, which characterizes the domain morphology and growth, 
scales as $C_{\psi\psi}(r,t) = g\textbf{(}r/\ell(t)\textbf{)}$ \cite{Binder}. Here, $g(x)$ is a scaling function independent of time. The average domain size $\ell(t)$ grows with time $t$ as
\begin{equation}
 \ell(t) \sim t^\alpha.
\end{equation}
\par
The growth exponent $\alpha$ depends upon the transport mechanism which drives segregation. For diffusive dynamics, $\ell\sim t^{1/3}$, which is referred to as the Lifshitz-Slyozov (LS) law \cite{Bray}. The LS behavior is the only growth law expected for phase-separating solid mixtures. However, for fluids and polymers, one expects faster growth at large length scales where hydrodynamic effects are dominant. For $d=3$, convective transport yields additional growth regimes \cite{Siggia} with
\begin{eqnarray}\label{alpha}
\alpha &=& 1,~~~~~\ell(t) \ll \ell_{\rm{in}},\nonumber\\
\alpha &=& 2/3,~~\ell(t) \gg \ell_{\rm{in}}.
\end{eqnarray}
In Eq. (\ref{alpha}), the inertial length $\ell_{\rm{in}}$ [$\simeq \eta^{2}/(\rho\gamma)$, $\eta$, $\rho$ and $\gamma$
 being the shear viscosity, density and interfacial tension] marks the crossover from a low-Reynolds-number viscous hydrodynamic regime to an inertial regime. There has been experimental evidence \cite{Chou} for a crossover from diffusive to viscous growth. However, no experimental observation of an inertial regime has been reported.
\par
While recent focus has turned to systems with realistic interactions and boundary conditions \cite{Das1,Blondiaux}, our understanding of segregation kinetics in bulk fluids remains far from complete. The viscous regime has been observed in numerical studies using the phenomenological Model H \cite{Puri, Shinozaki}. Further, both viscous and inertial regimes have been observed in lattice Boltzmann simulations \cite{Kendon1,Gonnella}. However, molecular dynamics (MD) methods, where hydrodynamics is automatically inbuilt, have  rarely been used to study domain growth, primarily due to heavy computational requirements. To the best of our knowledge, the first MD study was by Ma \textit{et al.} \cite{Ma}, who did not find a signature of viscous growth. In a later MD simulation, Laradji \textit{et al.} \cite{Laradji1} observed linear domain growth over a small interval in a binary Lennard-Jones (LJ) fluid. More recently, Thakre \textit{et al.} \cite{Thakre} used a similar model to study the crossover from diffusive to viscous dynamics. However, they do not observe linear growth in the post-crossover regime.  In all these cases, MD results have been obtained for low-density fluids over very limited time-windows, and conclusions drawn from these should not be taken seriously. In related work, Kabrede and Hentschke \cite{kabrede} found $\alpha\simeq0.5$ in MD simulations of gas-liquid phase separation. In this letter, we present results from large-scale MD simulations in conjunction with a numerical renormalization-group (RG) procedure \cite{Das3}. These state-of-the-art results will serve as a valuable reference for experimentalists and theorists in this area.
\par
Following Das \textit{et al.} \cite{Das2}, we have employed a symmetric model where particles of diameter $\sigma$ interact via the potential $V(r_{ij}) = U(r_{ij}) - U(r_{c}) - (r_{ij} - r_{c})dU(r_{ij})/dr_{ij}|_{r_{ij}=r_{c}}$. Here $U(r_{ij})=4\epsilon_{\alpha\beta}[(\sigma/r_{ij})^{12} - (\sigma/r_{ij})^{6}]$ is the LJ potential; and $r_{ij} = |\vec{r}_i-\vec{r}_j|$, $r_c = 2.5 \sigma$ and $\alpha$, $\beta =$ A, B. We chose $\epsilon_{\rm{AA}}=\epsilon_{\rm{BB}}=2\epsilon_{\rm{AB}} = \epsilon$, so that phase separation is favored energetically. All particles were assigned equal mass $m$. We set $m,~\sigma,~\epsilon$ and $k_B$ to unity. An incompressible fluid ($\rho=1$) is studied 
for which phase separation sets in at a critical temperature $T_c \simeq 1.423$ \cite{Das2}, well separated from the gas-liquid and 
liquid-solid transitions. A total number of $262144$ particles were confined in a cubic box of size $64^3$ with periodic boundary conditions in all directions. The MD runs were performed using the standard Verlet velocity algorithm 
\cite{Frenkel} with a time step $\Delta t = 0.01 \tau$, where the LJ time unit $\tau=(m\sigma^2/\epsilon)^{1/2}=1$. The temperature $T$ was controlled by a Nos\'{e}-Hoover thermostat (NHT) \cite{Frenkel}, which is known to preserve hydrodynamics. Homogeneous initial configurations were prepared by equilibrating the system at $T=10$. At $t=0$, the system is quenched to $T<T_c$. All the results presented here were obtained by averaging over $5$ independent runs at a quench temperature $T=0.77T_c$.
\begin{figure}[htb]
\centering
\includegraphics*[width=0.4\textwidth]{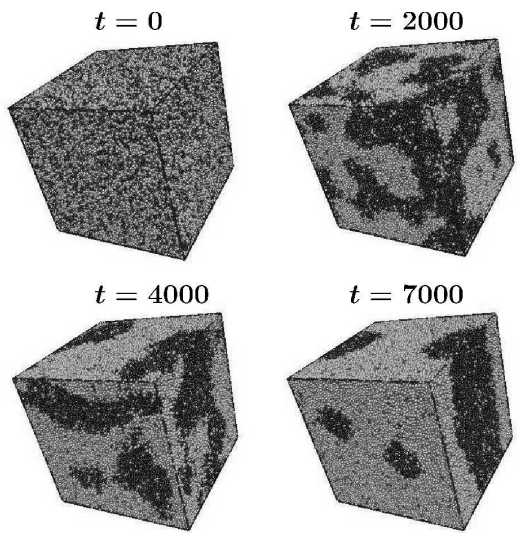}
\caption{\label{fig1} Evolution snapshots for a 50:50 binary Lennard-Jones fluid after quenching from the high-temperature homogeneous phase to $T=0.77T_c$. The A and B particles are marked in black and grey, respectively.}
\end{figure}
\par
In Fig. \ref{fig1}, we present evolution pictures at different times. As expected for a symmetric (critical) composition, a bicontinuous domain structure is seen. The snapshot at $t=0$ corresponds to the homogeneous state immediately after the quench. The snapshot at $t=7000$ corresponds to the situation where the system has almost completely phase-separated. Note that this time interval is more than an order of magnitude larger than earlier MD studies. While domains grow without encountering any perceptible size effects within this time-window, finite-size effects are seen beyond it \cite{SM,SA}.
\par
\begin{figure}[htb]
\centering
\includegraphics*[width=0.4\textwidth]{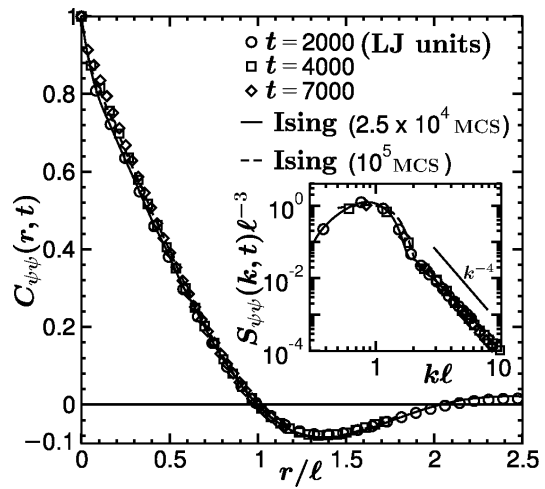}
\caption{\label{fig2} Scaling plot of the correlation function, $C_{\psi\psi}(r,t)$ vs. $r/\ell$, for $3$ different times. The inset shows the 
scaling plot of the structure factor $S_{\psi\psi}(k,t)$ for the same times. The solid and dashed lines denote analogous data from the Kawasaki-Ising model.}
\end{figure}

To characterize the domain morphology, we calculate the correlation function as
\begin{equation}\label{6}
C_{\psi\psi}(r,t) = \frac{\langle\psi(0,t)\psi(\vec{r},t)\rangle}{\langle\psi(\vec{r},t)^2\rangle},
\end{equation}
where the order parameter $\psi(\vec{r},t)~[=x_A(\vec{r},t)-x_B(\vec{r},t)]$ is the local concentration difference between A and B species. The angular brackets in Eq. (\ref{6}) denote statistical averaging. We use a coarse-graining (numerical RG) procedure \cite{Das3} to obtain the pure domain structure by eliminating thermal fluctuations in the snapshots of Fig. \ref{fig1}. Fig. \ref{fig2} shows the scaling plot of $C_{\psi\psi}(r,t)$ vs. $r/\ell$. The 
average domain size $\ell$ is defined as the first zero crossing of $C_{\psi\psi}(r,t)$, which is computed from the coarse-grained order parameter. Our correlation function data is comparable in quality to that obtained from a Langevin simulation \cite{Shinozaki}. The neat data collapse over an extended interval shows that a scaling regime is reached.
\par
We would like to make a 
quantitative comparison of our MD results for $C_{\psi\psi}(r,t)$ with those for segregation in the Kawasaki-Ising model (KIM), where the structure and 
dynamics are much better understood. The lines in Fig. \ref{fig2} denote $C_{\psi\psi}(r,t)$ vs. $r/\ell$, obtained from   a Monte Carlo (MC) simulation of the KIM \cite{Landau} with critical (50:50) composition. The MC scaling function is in excellent agreement with our MD data. In the inset of Fig. \ref{fig2}, we show the corresponding results for the scaled structure factor, $S_{\psi\psi}(k,t)\ell^{-3}$ vs. $k\ell$, where $S_{\psi\psi}(k,t)$ is the Fourier transform of $C_{\psi\psi}(r,t)$. Again, a good data collapse is obtained confirming the scaling
 form, $S_{\psi\psi}(k,t) = \ell^d f\textbf{(}k\ell\textbf{)}$ \cite{Bray}. The agreement with the corresponding KIM result is demonstrated again 
in the inset. Here, the decay of the tail with a power law, $S_{\psi\psi}\sim k^{-4}$, is consistent with the 
expected Porod's law \cite{Porod,BrayPuri}, $S(k,t) \sim k^{-(d+n)}$, for ordering dynamics in $d=3$ with a
 scalar order parameter ($n=1$). (In an extended publication, we will present results for morphological characteristics like the Tomita sum rule, Yeung-Furukawa law for $k\rightarrow0$, etc.) The excellent agreement of the scaling functions confirms a close similarity of structures 
formed during phase separation in fluids with those for solid mixtures. In a related context, Puri \textit{et al.} \cite{Puri2} have emphasized that domain growth morphologies are approximately independent of the kinetic mechanism of coarsening. This should be contrasted with a \textit{Cell Dynamical Systems} study by Shinozaki
 and Oono \cite{Shinozaki}, who argued that there were different scaling functions for phase separation in alloys and fluids.
\begin{figure}[htb]
\centering
\includegraphics*[width=0.4\textwidth]{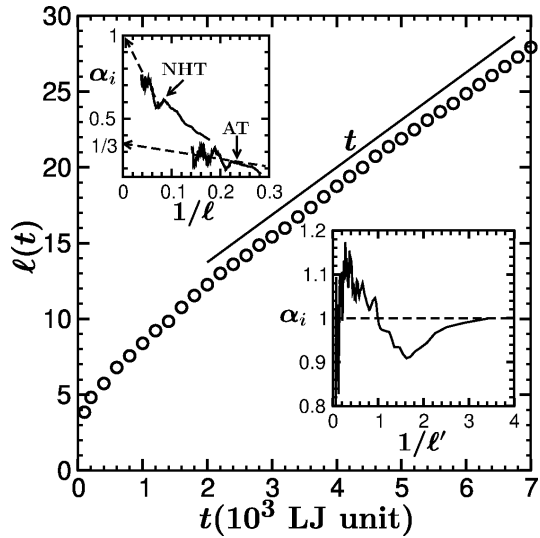}
\caption{\label{fig3} Plot of average domain size, $\ell(t)$, as a function of time $t$. The insets show the instantaneous exponent $\alpha_i$ vs. $1/\ell$ (upper) and $\alpha_i$ vs. $1/\ell^\prime$ (lower). A detailed explanation of the insets is provided in the text.}
\end{figure}
\par
Next, we focus on the time-dependence of the domain size. In Fig. \ref{fig3}, we plot $\ell$ vs. $t$. The growth at later times ($t>2000$) is clearly linear, but the earlier-time data deviates somewhat. In fact, a least-squares fit to the form $\ell(t)=B+At^\alpha$ in the range $t\in[0,2000]$ gives an exponent $\alpha\simeq0.75$. Essentially, the diffusive regime (with $\ell\sim t^{1/3}$) is very short-lived at this
 temperature. A gradual crossover to the viscous regime (with $\ell\sim t$) starts very early, extending over a large fraction of the time window, where the
 effective exponent is rather high.
\par
The top curve in the upper inset of Fig. \ref{fig3} shows the instantaneous exponent $\alpha_i[=\rm{d}(\rm{ln}\ell)/d(\rm{ln}t)]$ vs. $1/\ell$. Because of the significantly large off-set value in $\ell$ as $t\rightarrow{0}$, $\alpha_i$ tends to its asymptotic value in a linear fashion only in the limit $\ell\rightarrow\infty$ \cite{SM}. Without a knowledge of the functional form of $\alpha_i$ for the whole range of $1/\ell$, the dashed straight line serves as a guide to the eye and suggests $\alpha_i \rightarrow 1$. The lower curve in this inset is obtained from the coarsening dynamics when an Andersen thermostat (AT) is used instead of an NHT. The AT, where $T$ is controlled by letting the particles collide randomly with a heat bath, is stochastic in nature and does not model hydrodynamics. Thus, it is expected to provide 
a diffusive growth of domains with $\alpha=1/3$. Indeed our result is consistent with that expectation, as shown by the dashed straight line which extrapolates to $\alpha=1/3$ in the upper inset. This confirms the utility of the NHT in studying hydrodynamic phenomena in domain growth. We also emphasize that MD with an AT offers a more realistic way of modelling diffusive phase separation than the commonly-used but unphysical KIM.
\par
There is another instructive way of investigating $\alpha_i$. In hindsight, we introduce a time $t_0$ and assume that segregation kinetics follows a power-law behavior with time $t^\prime=t-t_0$:
\begin{equation}\label{ltat}
 \ell^\prime(t^\prime)=\ell(t)-\ell(t_0)=A{t^{\prime}}^\alpha.
\end{equation}
Then, we calculate the exponent $\alpha_i=d(\rm{ln}\ell^\prime)/d(\rm{ln}t^\prime)$. For linear growth,
 Eq. (\ref{ltat}) is invariant under an arbitrary choice of $t_0$. Thus, if $t_0$ is chosen appropriately, $\alpha_i\simeq 1$ for all values of $t^\prime$.  However, as noted by other authors \cite{Shinozaki,SM}, in computer simulations of finite systems one finds an oscillation of $\alpha_i$ as $\ell\rightarrow \infty$, with growing amplitude around the expected value. This is due to increasing separation between the domains of like particles, thus delaying collisions between domains of \textit{large} size. The lower inset of Fig. \ref{fig3} plots $\alpha_i$ vs. $1/\ell^\prime$ for $t_0=2500$, which lies in the linear region. Indeed, this plot is consistent with the above expectation, and $\alpha_i$ oscillates around the mean value $\alpha=1$.
\begin{figure}[htb]
\centering
\includegraphics*[width=0.4\textwidth]{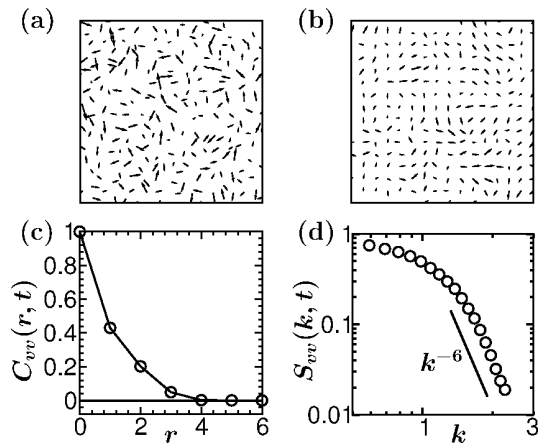}
\caption{\label{fig4} Pattern formation in the velocity field at $t=7000$. (a) Projection of velocity field onto a $2$-$d$ cross-section. (b) Same as (a) but for the coarse-grained velocity field. (c) Plot of correlation function, $C_{vv}(r,t)$ vs. $r$, for the velocity field. (d) Plot of corresponding structure factor, $S_{vv}(k,t)$ vs. $k$, on a log-log scale.}
\end{figure}
\par
Finally, we turn our attention to the pattern dynamics of the velocity field. Fig. \ref{fig4}(a) shows a $2$-$d$ cross-section of the system at $t=7000$, with particle velocities being projected onto this plane. While the orientations of velocity vectors look fairly random at the microscopic scale, structure starts emerging upon coarse-graining over larger length scales, as seen in Fig. \ref{fig4}(b). In Figs. \ref{fig4}(c) and \ref{fig4}(d), we present plots of the velocity correlation function $C_{vv}(r,t)$ and structure factor $S_{vv}(k,t)$, analogous to the density field. The decay of the structure factor tail as $S_{vv}\sim k^{-6}$ is consistent with the generalized Porod's law \cite{BrayPuri} for ordering of a $3$-component ($n=3$) vector field in $d=3$. This is indicative of pattern formation with monopole-like defects, which can be seen in the coarse-grained snapshots of the velocity field. To study growth in this ordering, as seen in the lattice Boltzmann simulations \cite{Gonnella,Kendon1}, we need coarse-graining over a larger length scale which is not accessible with the present system size.
\par
In summary, our results from extensive MD simulations of a binary LJ fluid 
unambiguously confirm the linear growth law in the viscous hydrodynamic regime. We use a numerical RG technique to obtain noise-free data from our MD studies.
Even though the growth mechanisms in fluids are different from solids, the domain morphologies are comparable in 
the two cases. Similar studies at higher quench temperatures will be useful to identify 
 diffusion-driven growth and crossovers at early times. Note that while our choice of system size was appropriate for understanding viscous growth, this size is not large enough to study 
dynamics in the inertial regime. To achieve the latter goal, sophisticated modelling at the multi-scale level is required, in addition to usage of parallel programming and a graphics card.
\par
SKD acknowledges useful discussions with M. Laradji. SA thanks University Grants Commision, India, for providing a fellowship and Jawaharlal Nehru Centre for Advanced Scientific Research, Bangalore, India, for supporting her visits.\\
$^*$das@jncasr.ac.in

\end{document}